\documentclass[draft]{agujournal2019}
\usepackage{graphicx}
\usepackage{url} 
\usepackage[inline]{trackchanges} 
\usepackage[percent]{overpic}
\usepackage[section]{placeins}
\usepackage{pbox}
\usepackage{comment}
\usepackage{titling}
\usepackage{natbib}
\usepackage{soul}
\usepackage{bm}
\usepackage{siunitx}
\usepackage{textcomp}
\newcommand{\GG}[1]{}

\usepackage{ragged2e}
\usepackage[colorlinks=true, urlcolor=blue]{hyperref}  
\usepackage{hyperref}
\usepackage{orcidlink}
\usepackage{subcaption}
\justifying

\journalname{Space Weather}

\begin{document}

\title{The effect of uncertainties in reproducing the ambient solar wind at Earth on forecasting CME arrival times}

\authors{
    Syed A. Z. Raza\affil{1},  
    Talwinder Singh\affil{2},  
    Nikolai V. Pogorelov\affil{1,3}
}

\affiliation{1}{Department of Space Science, The University of Alabama in Huntsville, AL 35805, USA}
\affiliation{2}{Department of Physics \& Astronomy, Georgia State University, Atlanta, GA 30303, USA}
\affiliation{3}{Center for Space Plasma and Aeronomic Research, The University of Alabama in Huntsville, AL 35805, USA}

\correspondingauthor{Syed A. Z. Raza}{sar0033@uah.edu}

\begin{abstract}

Coronal Mass Ejections (CMEs) are the major drivers of Space Weather (SWx), so predicting their arrival at Earth is a major aspect of SWx forecasting. Despite increasingly complex models proposed over the past decades, the mean absolute error (MAE) for predictions of CME arrival still surpasses 10 hours. In this study, we use machine learning (ML) techniques trained on the discrepancies between observed and modeled solar wind (SW) at the L1 point, upstream of CMEs, to quantify and ''correct'' the errors in CME Time of Arrival (TOA) associated with these discrepancies. We use CME data from the Database Of Notifications, Knowledge, Information (DONKI) developed by the NASA Community Coordinated Modeling Center (CCMC) for our investigation. The WSA-ENLIL-Cone (WEC) model inputs and outputs are available on DONKI for each CME, along with the associated forecast errors. The dataset consists of 122 CME events observed between March 2012 and March 2023. SW properties at L1 and publicly available simulation results based on the WEC model are obtained. Three machine learning (ML) models are employed: 1) k-nearest neighbors (KNN), 2) support vector machine (SVM), and 3) linear regression (LR). Univariate and multivariate ML schemes were developed to examine how individual features and their combinations contribute to reducing the MAE in CME TOA forecasts. The best univariate and multivariate models improved the forecast by 36.6 and 45 minutes, respectively. 

\section*{Plain Language Summary}

Coronal Mass Ejections (CMEs) are massive explosions in the Sun's lower atmosphere. Many CMEs are Earth-directed and can give rise to Space Weather (SWx) effects in the near-Earth environment. These effects can cause damage to satellites and pose risks to humans outside the protection of Earth's magnetic field. Consequently, many physics-based and empirical models have been developed to predict CME time of arrival (TOA) and take necessary precautions to mitigate their effects. However, the mean absolute error (MAE) of CME TOA predictions remains over 10 hours, motivating an evaluation of the sources of these errors. One such error source is inaccuracies in the ambient solar wind (SW) representation within the models. CMEs travel through the interplanetary medium, interacting with the ambient SW. The main goal of this paper is to quantify the extent to which errors in SW modeling affect CME TOA predictions on Earth. The inputs for the machine learning models are the differences between modeled and observed SW parameters. We found that correcting for SW errors can reduce the the MAE by $\approx$ 1 hour.

\end{abstract}

\begin{keypoints}
\item WSA-Enlil-Cone model results in $\approx$ 12 hours of mean absolute error.
\item The error in SW properties can be used to reduce MAE by $\approx$ 1 hour.
\item Errors in CME initial properties contribute most to the MAE.
\end{keypoints}

\section{Introduction}

Coronal mass ejections (CMEs) are powerful plasma eruptions emerging from the lower corona. They are constrained by twisted magnetic fields and are known to be among the most violent events in our solar system \citep{Chen11}. CMEs are frequently directed towards Earth. They travel through the interplanetary medium and inject enormous amount of energy into the Earth's upper atmosphere, notably the magnetosphere, causing so-called geomagnetic storms \citep[e.g.][]{Zhang2007, Chi2016, Sharma2013}. These storms can have severe consequences for human health, and space-borne and ground-based instruments. They pose a major threat to  telecommunications, airlines, navigation (e.g., global positioning system), power grids, etc. \citep[e.g.][]{Pulkkinen07, Millward13}. It is therefore crucial to identify Earth-aimed CMEs, predict their properties and time of arrival (TOA) at 
Earth.

To address the scientific challenge of forecasting CME TOA at Earth, several propagation models have been developed. The WSA-ENLIL+Cone (WEC) model is a widely used framework to simulate large-scale CMEs as they evolve up to 1 AU. Space agencies including National Oceanic and Atmospheric Administration (NOAA), UK Met Office, and many others use the WEC model for Space Weather (SWx) predictions. WEC uses the following three components: (1) The Wang--Sheeley--Arge \citep[WSA][]{Arge00, Arge04} model is a semi-empirical coronal model which predicts SW conditions within the spatial domain of 21.5 $R_{\odot}$, (2) ENLIL 
\citep{Odstrcil04, Odstrcil99a, Odstrcil99b, Odstrcil2003} is an MHD model of the inner heliosphere (covering distances from 21.5 $R_{\odot}$ outwards), and (3) the Cone model \citep[]{Xie2004, Xie2006} is used to determine the CME geometry from remote observations. CMEs with parameters provided by the Cone model are inserted into the SW at the inner boundary at 21.5 $R_{\odot}$ (for ENLIL) and are allowed to expand. 
Several other models are also used, including empirical models \citep{Vandas1996, Brueckner1998, Gopalswamy01, Gopalswamy05, Wang2002, Manoharan2004}, drag-based models \citep{Vrsnak01, Vrsnak02}, and  physics-based models, e.g., the Shock Time of Arrival \citep[STOA,][]{Dryer1984} and STOA-2 \citep{Moon2002}. \citet{Zhao&Dryer2014} performed a comprehensive study comparing the capabilities of these models. However, despite a variety of numerous approaches listed above, \cite{Riley2018} showed that the mean absolute error (MAE) for CME TOAs still surpasses 10 hours. In a recent paper, \cite{Kay24} considered
a larger dataset involving a number of recent CME events
and came to conclusions similar to those of \cite{Riley2018}.

\citet{Riley2018} also pointed out that escalating the model complexity does not necessarily lead to improvements in CME TOA esitimates. Several studies  highlighted potential sources of this persisting error \citep[see, e.g.,][]{OwensandCargill2004, Mays2020, Riley&BenNun2021}. Two of the primary causes are (a) uncertainty in the estimated CME parameters, i.e., speed, angular width, direction, and overall geometrical structure near the Sun, and (b) the forces that act upon CMEs in the inner heliosphere due to the CME-SW interaction. Earlier studies \citep[see, e.g.,][]{Kay15, Jones2017} showed that CMEs tend to deflect towards the 
heliospheric current sheet (HCS),  most of these deflections occurring close to the 
Sun at heliocentric distances below 4 $R_{\odot}$). \citet{Sahade2020} performed MHD simulations of an isolated CME flux rope in the presence of a Coronal Hole (CH) with the goal of understanding the effect of various properties of CHs (such as magnetic field, area, distance) on CME deflection. They found that CME deflection is directly proportional to the CH magnetic field strength and area.
According to \citet{Gui2011}, such deflections are primarily driven by variations in the magnetic energy density in the corona, which make CMEs shift towards areas of lower magnetic pressure. In addition to the coronal magnetic field, CMEs are also affected by the ambient SW. They can be accelerated or decelerated by fast or slow SW streams \citep{TuckerHood2015}. \citet{Savani2010} studied the distortion of the 2007, November 14 CME. The leading edge of this CME (observed from STEREO-B field-of-view) was distorted from a circular to an increasingly concave structure due to its interaction with the bimodal solar wind. The SW generated by MHD models can also be inaccurate. 
\citet{Owens2008} showed that the WEC model underestimated all three magnetic field vector components at L1 and resulted in  a much lower variability in observed SW speed. They also showed that WEC dropped its predictive capabilities at the solar maximum, when the CME frequency becomes high and HCS inclination starts to increase. Inaccuracies in the SW properties at L1 are also associated with the magnetograms used to derive inner boundary 
conditions \citep[see, e.g.,][]{Riley2018}).
\citet{Shen2012} investigated in detail the acceleration and deceleration dynamics of CMEs and the forces acting upon them. CMEs can also collide with each other, which can result in their merging and acceleration/deceleration \citep[e.g.,][]{Singh2020, Mishra2016, Shen2017}.

In recent years, ML methods are becoming increasingly popular for forecasting CME TOA. One category of ML models includes those that are purely data-driven, without any physics incorporated. 
A general workflow for these models is as follows. 

(1) A CME database is constructed for Earth-directed CMEs, and their onset and TOAs are collected (from such sources as the Richardson and Cane catalog or the CME scoreboard at the NASA's Community Coordinated Modeling Center (CCMC)). 

(2) The feature space is built for these ML models using estimates for the physical properties of CMEs, i.e., angular width, mass, initial speed at 21.5 $R_{\odot}$, source region, etc. These properties can be obtained from the SOHO LASCO CME Catalog by matching the CME onset time. In situ SW properties are also often used as features (inputs). These include such physical quantities as density, temperature, total pressure, bulk speed, and all three components of the magnetic field.
\citet{Liu2018CATPUMA} developed an engine called CME TOA Predictions Using Machine learning Algorithms (CAT-PUMA). CAT-PUMA is based on Support Vector Machine (SVM) algorithms. They used CME and SW properties (6 hr averaging after a CME onset) 
as inputs (features) for their model, with the output (target) being the CME transit (or arrival) time. Their dataset comprised 182 partial- and full-halo CME events spanning two decades (1996-2015), which were split into $80\%$ training and $20\%$ testing sets, and the model was trained on 145 events and then tested on the 37 event. For the transit time predictions of the test set, CAT-PUMA's 
model the TOA MAE turned out to be $\sim 5.9$ hours. \citet{Wang2019} employed a convolutional neural network (CNN) for TOA predictions. Their model only required white-light SOHO LASCO C2 images as inputs, and utilized a dataset consisting of 223 geoeffective CMEs with the total of 1122 images. They reported MAE of \(\sim 12.4\) hours which is comparable with the number obtained with the physics-based models described in the literature. \citet{Alobaid2022} extended both of the above-mentioned studies by adopting an ensemble learning approach. They used CME properties, SW properties, and white-light images as input. They developed several regression models and a CNN model, as their input data requires both numerical and image processing. The results of several ML models were further combined to derive an ensemble output. This methodology was applied to CME events from solar cycles 23 and 24, and they obtained an MAE of 9.75 hours. \citet{Chierichini2024} further extended the work of \citet{Liu2018CATPUMA} by studying the strengths and weaknesses of CAT-PUMA. In addition to TOA predictions, \citet{Chierichini2024} and \citet{Fu2021} also delved into forecasting the geoefffectiveness of CMEs using ML methods. 

The main idea of this paper is inspired by the UK Met Office's ``human-in-the-loop" application to the WEC model outputs \citep{Riley2018}.  It allows a SWx forecaster to ``tweak'' the raw output of the WEC model before the CME arrival depending on the MHD SW predictions. 
For example, if the ENLIL-predicted background SW speed at 1 AU is higher than it was observed in situ, the CME TOA can be adjusted so that it arrives later than the corresponding CME obtained from the raw WEC model.  Our main goal is to automate this process using ML and reduce the WEC error associated with the simulated background SW. As a result of our proposed ML analysis, we found that CME TOA error can be improved by 5-6\%.

The structure of the paper is as follows: Section \ref{data-and-methods} describes the CME, observed SW, and simulated SW data used in our analysis. Section \ref{result} shares some of the main scientific findings of our research. Lastly, section \ref{discussion} provides a discussion of the obtained results. 


\section{Data and Methods}
\label{data-and-methods}
\subsection{CME data mining}

To build a database suitable for ML analysis, our first step was to compile a list of CMEs that not only arrived at Earth and caused geomagnetic disturbances, but also identify those for which the ambient SW data was available. The Space Weather Database of Notifications, Knowledge, Information (DONKI) \citep{Maddox2014} is a one-stop online tool for SWx researchers and forecasters. It has been developed at the Community Coordinated Modeling Center (CCMC), and we use it to collect CME data. The DONKI database has a search functionality to facilitate data mining and can be accessed online at \url{https://kauai.ccmc.gsfc.nasa.gov/DONKI/}. DONKI includes the WEC model input and output data, which can be accessed using its user-friendly web service API (introduced at the link above), making it a suitable resource for building an inventory of CME events for our study. We used the DONKI data for the time period of 11 years, from March 2012 to March 2023. The justification for choosing this time period is provided in section \ref{omni-enlil}. Below, we summarize the key CME parameters provided on the DONKI webpage.

\begin{enumerate}
    \item The CME Cone properties, i.e., speed, angular width, longitude, and latitude. These properties are required to reconstruct CMEs at ENLIL model inner boundary at 21.5 $R_{\odot}$. The estimated TOA at $R=21.5\,R_{\odot}$ for each CME is also reported. This is an important parameter for our study as it approximates the time of CME departure from the solar corona. The measuring instruments for each event, e.g., LASCO/C2 or C3 aboard SOHO, or STEREO A: SECCHI/COR2 etc, are also listed. 
    
    \item The WEC model output, i.e, CME observed and shock TOA at Earth, duration of a disturbance in hours, the minimum magnetopause stand-off distance, $K_{p}$ indices, along with arrival at other locations (e.g, STEREO A or B, Lucy, PSP, etc.), if any occurred.
     
\end{enumerate}

\begin{figure}[h!]
    \centering    \includegraphics[width=0.8\textwidth]{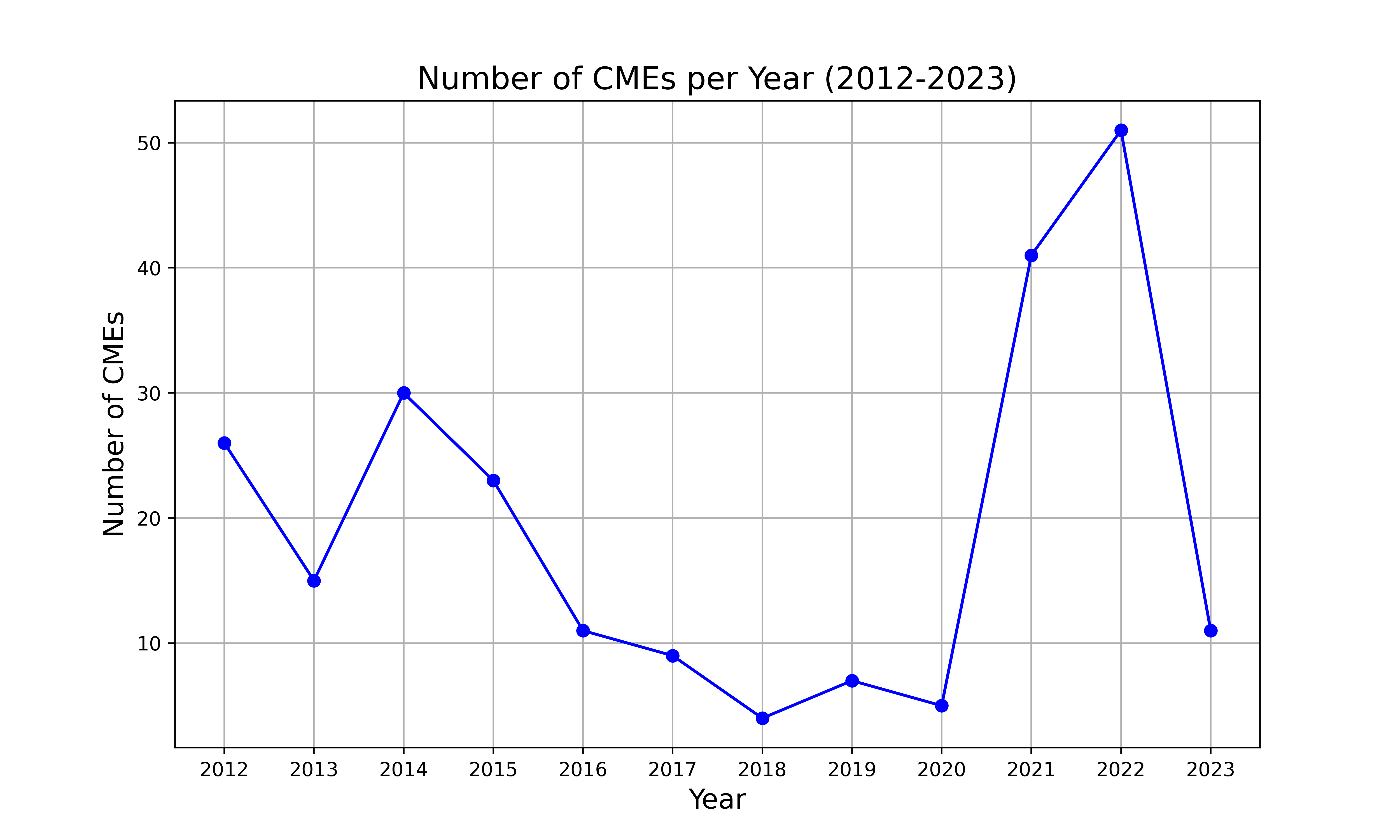}
    \caption{Number of Earth-directed CMEs per Year (2012-2023) reported in DONKI.}
    \label{fig:nCMEs}
\end{figure}

DONKI categorizes CME arrivals and interplanetary shocks (IPS) under a single identification code, making all arrival entries labeled as ``Earth Shock Arrival Time". These entries encompass either CME shocks or CME arrivals that did not drive shocks but still impacted Earth. For such events, an IPS signature is added if a possible signature of flux rope in the magnetic field distribution is observed. redAs an example, the results for the January 30, 2014 CME can be accessed at  \url{https://kauai.ccmc.gsfc.nasa.gov/DONKI/view/WSA-ENLIL/4541/1}. The most important parameters for our study were the CME shock TOA, because it is related to the CME TOA, time at the injection height (21.5 $R_{\odot}$), when a 
CME starts to interact with the ambient medium, and the CME Cone model parameters. For the 11 year period under consideration, these parameters were extracted manually from the webpage for a total of 233 available, Earth-directed CMEs. Figure \ref{fig:nCMEs} shows the number of CMEs per year for the chosen time interval. The year of 2018 had the least number of CMEs listed (only 4), whereas 2021 and 2022 had the greatest number of CMEs, 41 and 51, respectively. The time interval over the second half of 2019 corresponds to the solar minimum, as can be seen from the low number of events. After 2020, solar activity picks up as can be seen in 2021 and 2022. The year of 2023  shows only 11 CME events because the data only contains events that occurred till March.

\subsection{Data pre-processing}

It is important to note that DONKI cannot report WEC CME forecast errors when multiple CMEs are inserted into the same background SW simulation. In such cases, DONKI reports the forecast error only for the earliest arriving CME. For instance, Table \ref{sameCMEs} shows a collection of three CMEs from November 2013 that were inserted into the same background. 
The prediction error reported for all three events is the same and corresponds to the earliest arriving CME. However, it remains unknown which of the three CMEs arrived first. Thus, it is unclear which of the Cone model parameters resulted in PE \( = -11.8\), making it impossible to reproduce the in situ SW with a reasonable degree of certainty. All such events were deleted from our data. In the end, we obtained 122 CMEs in total that did not include duplicates. The upstream SW data was available for all of them (more about this in section \ref{omni-enlil}). 

\begin{table}[t]
\centering
\setlength{\tabcolsep}{6pt} 
\renewcommand{\arraystretch}{1.2} 
\resizebox{\textwidth}{!}{ 
\begin{tabular}{|c|c|c|c|c|c|c|}
\hline
Event \# & Longitude & Latitude & Speed (km/s) & Half-Angle & Time 21.5 & PE (hours) \\ \hline
CME1 & -10.0 & -11.0 & 530.0 & 27.0 & 2013-11-07T22:02Z & -11.8 \\ \hline
CME2 & -6.0  & -21.0 & 444.0 & 18.0 & 2013-11-08T12:26Z & -11.8 \\ \hline
CME3 & -9.0  & -29.0 & 508.0 & 27.0 & 2013-11-08T15:33Z & -11.8 \\ \hline
\end{tabular}
}
\caption{CME characteristics including longitude, latitude, speed, half-angle, time 21.5 values, and prediction error (PE) for three closely occuring events.}
\label{sameCMEs}
\end{table}

For the remaining 122 CMEs, after data cleaning, the WEC forecast error can be calculated by taking the difference between the predicted and observed shock TOA.
\begin{equation}
\Delta t = t_\mathrm{predicted} - t_\mathrm{observed} 
\label{eq:01}
\end{equation}

\begin{figure}[t]
    \centering
    \includegraphics[width=\textwidth]{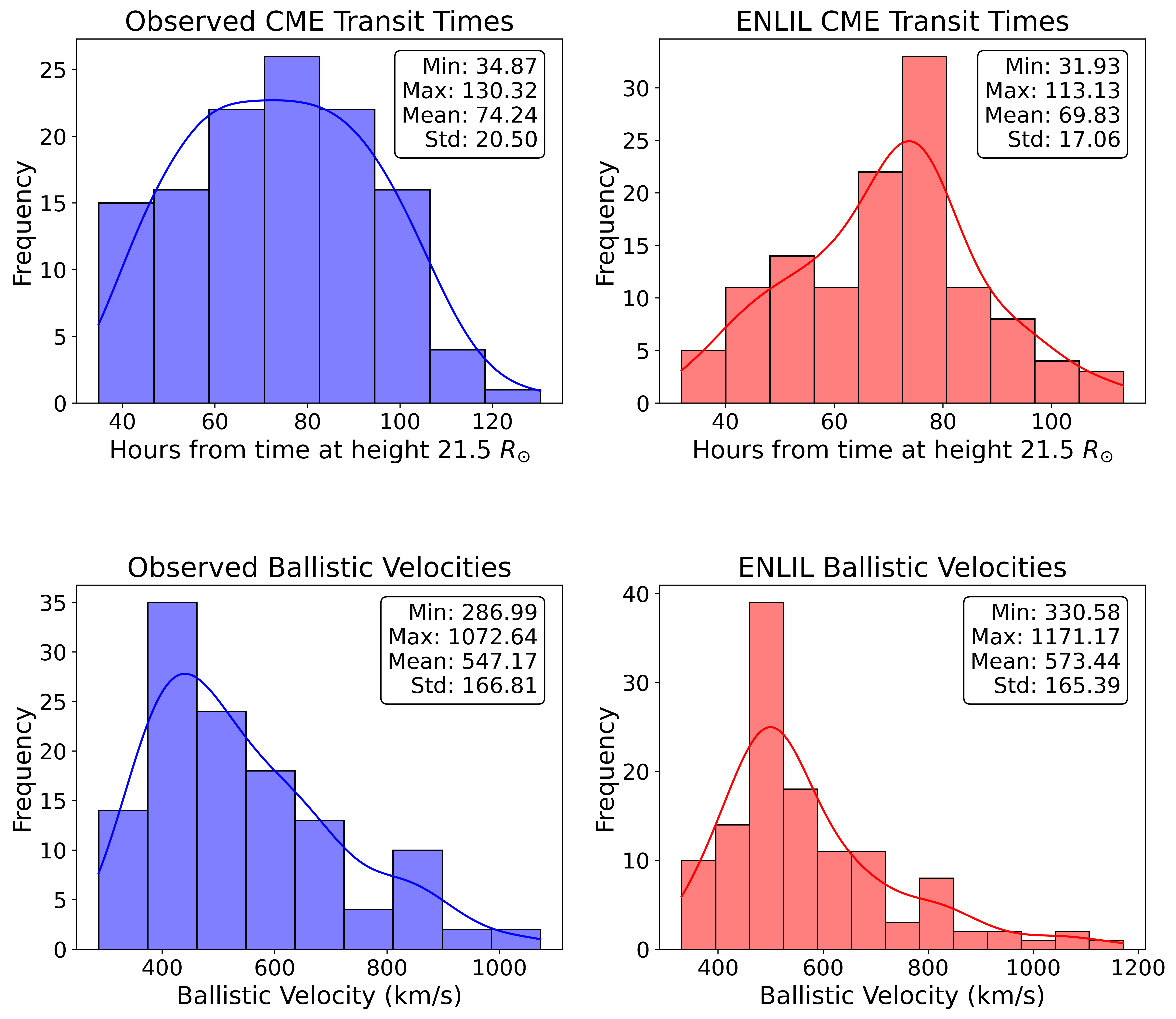}
    \caption{Comparison of the observed and ENLIL transit distribution (top row) and ballistic velocity distribution (bottom row). Transit times are given in hours between the CME insertion at height 21.5 $R_{\odot}$ and Earth arrival. Ballistic velocities are given in km/s. The maximum, minimum, mean, and standard deviations for all groups are also shown.}
    \label{fig:transit_comparison}
\end{figure}

Negative $\Delta t$ indicate early TOA predictions.
Positive $\Delta t$ suggests that a CME was observed earlier than the predicted shock TOA. The top two panels Figure \ref{fig:transit_comparison} shows the CME transit time, from $R=21.5 \,R_{\odot}$ to the Earth, distribution for both the observed (left panel) and 
WEC-simulated (right panel) CMEs. The minimum, maximum, and mean transit times are also shown. It took a little more than 34 hours for fastest  observed CMEs to reach Earth, whereas the slowest one arrived in about 5 days 10 hours. The bottom two panels of Fig.~\ref{fig:transit_comparison} show the ballistic speeds of the observed (left) and simulated CMEs (right). We estimate the ballistic speed as \( v_{\text{ballistic}} = \frac{0.9\,\text{AU}}{\text{transit time}} \), assuming constant-speed propagation between 0.1 AU and 1 AU. The mean error (ME) for CME TOAs was calculated as
\begin{equation}
\text{ME} = \frac{1}{N}\sum\limits_{i=1}^{N} \Delta t_{i} 
\label{eq:02}
\end{equation}

It turned out to be -4.42 hours (4 hours 25 minutes), which indicates a bias towards early predictions. Although the ME can reveal biases in prediction models, much information is lost as the $\Delta t$'s with opposite signs cancel  out each other. MAE is a more appropriate metric to address this limitation, because it preserves the magnitude of time differences between the observations and forecasts by averaging the absolute values of forecast errors, irrespective of their sign\citep[see, e.g.,][]{morley2018measures}.

\begin{equation}
\text{MAE} = \frac{1}{N} \sum_{i=1}^{N} |\Delta t_{i}| 
\label{eq:03}
\end{equation}

For the chosen 122 events, the MAE is 12.31 hours, which indicates that, on average, the model prediction is 12.31 hours off, but can be in either direction. The MAE for our dataset is comparable to other studies in the literature. The standard deviations for the observational and simulation results are 20.50 and 17.06, respectively, suggesting that the observed transit times tend to have a wider spread around the mean and are more dispersed. Same trend can be seen in the ballistic speeds (observations are more dispersed than simulations). In contrast, WEC-predicted transit times have lower variability and are more tightly clustered around the mean.

\subsection{Upstream Solar Wind data}
\label{omni-enlil}

Observed SW parameters; i.e, bulk flow speed, number density, temperature, and the three Radial-Tangential-Normal (RTN) components of the magnetic field vector were obtained from OMNI data \citep{Papitashvili2014} for the 11 year interval under investigation, with a temporal resolution of 1 hour. Magnetic field magnitude is then calculated from its three components: $B_{\text{mag}} = \sqrt{B_{\text{radial}}^2 + B_{\text{tangent}}^2 + B_{\text{normal}}^2}$. \citet{Singh18} have shown that total SW pressure plays an important role in CME dynamics. It is calculated with the following formula (in CGS units): 
\begin{equation}
P_{\mathrm{total}} = 2n k T + \frac{B_{\mathrm{mag}}^{2}}{8 \pi} \label{eq:04}
\end{equation}

The factor of 2, introduced with the thermal pressure term, comes from the assumption that electron and ion temperature are approximately equal in the SW state at 1 AU ($T_{i} = T_{e} =T$). Additionally, to obtain the simulated SW profile ahead of a CMEs, WSA-ENLIL SW data files were sourced from the Community Coordinated Modeling Center (CCMC). These files are publicly available online at \texttt{\url{https://iswa.ccmc.gsfc.nasa.gov/iswa_data_tree/model/heliosphere/wsa-enlil-Cone/velocity-density-timeline-DATA/}}. Each file represents the SW environment into which a CME is inserted, with filenames corresponding to date and time of each CME listed in the DONKI database. As observed from the above link, simulated SW data have been consistently available since March 2012, which marks the start time of our study. These filenames were retrieved from the DONKI webpage, ensuring a comprehensive data collection for the upstream SW profiles for each of the 122 CMEs included in our study.

For each CME event, we monitored the SW data at the L1 point from the time of CME insertion in the ENLIL simulation at 21.5 $R_{\odot}$ to its TOA at Earth. As mentioned above, the time at a height of 21.5 $R_{\odot}$ is available for each CME event in the DONKI database. This time instant marks the beginning of the CME-SW interaction in the ENLIL domain. The end time of our analysis is chosen as the earliest between the observed or simulated TOAs. This is done for each CME, ensuring that our analysis only includes the time interval before CME arrival. For each CME event, the difference between the observed and simulated SW parameters is calculated over the chosen time interval. Then, the mean of this difference is calculated to represent the SW discrepancies at L1 point with a single value per parameter per CME event.

\subsection{Machine learning models}


We used several ML models to improve the prediction accuracy of CME TOA at Earth. When treating an ML model as a black box, features are the parameters the model takes as input and the target is the variable the model predicts. For our ML procedure, features are the $\Delta E_{sw}$, where E represents a SW property (it could be magnetic field, flow velocity, temperature, density, or total pressure), and $\Delta$ represents the mean of differences between observed and simulated values of $E$ from the time of CME insertion to its TOA. Cone model parameters obtained from DONKI are also used as features. These include the CME speed, angular half-width, and propagation direction, represented by the longitude and latitude of the apex of its assumed conical shape. The target variable, the one that ML model predicts is the CME prediction error. 
One of the most important aspects of an ML procedure is cross-validation. During cross-validation, the data is resampled into training and testing sets. The ML model is trained on the training set and the testing set is used to validate the model's ability to make predictions on an independent dataset.

\subsubsection{K-Nearest Neighbors (KNN)}
K-nearest neighbors (KNN) is a non-parametric, supervised ML algorithm which can be used for regression analysis.
We use the Euclidean distance as a measure of the "nearness." Once the nearest neighbors are identified, a prediction is made using the average of the observed values at those points. Mathematically, for a new input vector \( \mathbf{x}_0 \), the predicted output \( \hat{y}(\mathbf{x}_0) \) is given by
\begin{equation}
\hat{y}(\mathbf{x}_0) = \frac{1}{k} \sum_{i \in \mathcal{N}_k(\mathbf{x}_0)} y_i
\end{equation}

Here \( \mathcal{N}_k(\mathbf{x}_0) \) denotes the set of the \( k \) nearest neighbors of \( \mathbf{x}_0 \) in the training dataset, and \( y_i \) represents the target value corresponding to the \( i \)-th neighbor. This equation describes a uniform averaging approach. Alternatively, a weighted average can be calculated, where higher importance is assigned to the points in \( \mathcal{N}_k(\mathbf{x}_0) \) that are closer to \( \mathbf{x}_0 \). 

The performance of a K-Nearest Neighbors (KNN) regressor is highly dependent on the choice of model hyperparameters. The hyperparameters used in our implementation of KNN are: (1) The number of neighbors k, (2) the weighting scheme for the neighbors (either uniform or distance-based weighting), and (3) the algorithm used to compute the nearest neighbors (we used auto, ball tree, kd tree, and brute algorithms for this). We employ Python's \texttt{GridSearchCV} to perform an exhaustive search over the defined hyperparameter grid, employing a 5-fold cross-validation to estimate the performance of each parameter combination. The best-performing instance of the KNN model, as determined by the grid search, is then used to predict the target values for the test dataset.

\subsubsection{Linear Regression}
Linear Regression (LR) is a supervised parametric ML method that assumes a linear relationship between the features and the target variable. Mathematically, for multiple features, the relationship is expressed by the following equation:
\begin{equation}
y = \beta_0 + \beta_1x_1 + \ldots + \beta_nx_n + \epsilon.
\label{eq:07}
\end{equation}

In this equation, \( y \) is the target variable, \( x_1, x_2, \ldots, x_n \) are the input features, \( \beta_0 \) is the intercept, \( \beta_1 \), \( \beta_2 \), \ldots, \( \beta_n \) are the coefficients (weights) associated with each feature, and \( \epsilon \) is the error term representing the model residuals. The goal of Linear Regression is to estimate the coefficients \( \beta_0 \), \( \beta_1 \), \ldots, \( \beta_n \) so that the sum of the squared differences between the observations and model predictions is minimized. This process is known as Ordinary Least Squares (OLS). Unlike models that require hyperparameter tuning (such as K-Nearest Neighbors), the Linear Regression performance is primarily evaluated depending on how well the linear assumptions hold for the data and how well the model generalizes to an independent dataset.

This process is easy to visualize in two-dimensions (2D). Suppose there is only one feature (\( x_1 \)) and one target (\( y \)). Equation \ref{eq:07} simplifies to
\begin{equation}
y = \beta_0 + \beta_1x_1,
\end{equation}

This is simply the equation of a straight line in the $x_1$-$y$ plane, where \( \beta_0 \) represents the y-intercept and \( \beta_1 \) is the slope. During training, the model estimates \( \beta_0 \) and \( \beta_1 \) so that the sum of the squared vertical distances from the data points to the resulting line is minimized. The model then uses this line to find y-values (estimates) for 
each $x_{1}$-value in the testing set. 

\subsubsection{Support Vector Machine (SVM)} 

Support Vector Machine (SVM) is a non-parametric ML method that does not assume any relationship between the features and the target. The primary objective of SVM is to find a hyperplane with the best to the data while maintaining a margin of tolerance for errors. This makes it extremely effective in high-dimensional feature spaces.

The following three hyperparameters control the model complexity and accuracy: (1) the penalty parameter C, which regulates the trade\text{-}off between the margin width and the model complexity,  (2) the kernel function (linear or radial basis function) looks for a hyperplane that fits the data, and (3) the kernel coefficient $\gamma$, which defines how far the influence of a single training example extends. 

The hyperparameter optimization is performed in way similar to the KNN model, using a hyperparameter grid involving \( C \), \( \gamma \), and the kernel. Python's \texttt{GridSearchCV} further performs a 5-fold cross-validation on this hyperparameter grid, and the best-performing combination of hyperparameters is selected.

\section{Results}
\label{result}

\subsection{PE correlations with SW and Cone model parameters}

\begin{figure}[t]
    \centering    \includegraphics[width=\textwidth]{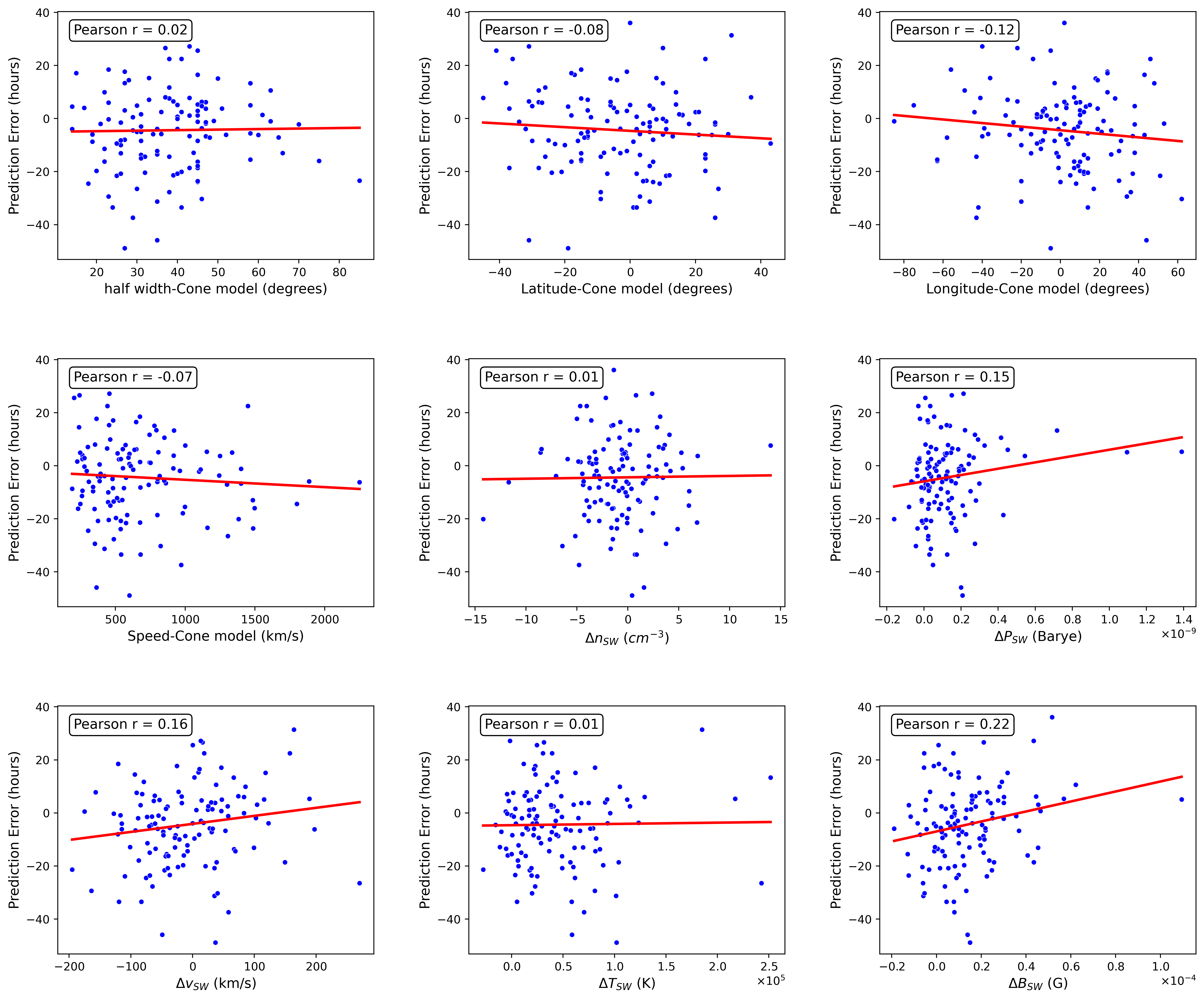}
    \caption{This figure illustrates the linear relationship between WEC forecast error  and Cone model parameters: half-width, latitude, longitude, and CME speed. It also linearly compares the WEC forecast error with the upstream SW difference parameters; $\Delta n_{SW}$, $\Delta P_{SW}$, $\Delta v_{SW}$, $\Delta T_{SW}$, and $\Delta B_{SW}$}
    \label{fig:correlation_plots}
\end{figure}

The feature space consists of a total of nine parameters. These include discrepancies in SW pressure (\( \Delta P_{SW} \)), speed (\( \Delta v_{SW} \)), magnetic field (\( \Delta B_{SW} \)), temperature (\( \Delta T_{SW} \)), and density (\( \Delta n_{SW} \)), and the four Cone model parameters CME initial speed, CME half width, along with longitude and latitude of CME apex. Figure \ref{fig:correlation_plots} shows the scatter plots of the PE with all the features; 5 SW parameters mentioned above and the 4 Cone model parameters. The Pearson correlation coeffecients are also shown. It is observed that all 9 features exhibit weak to no linear correlation with the PE.

\subsection{Machine Learning}

\subsubsection{Leave one out cross-validation}

For our ML task, the 9 parameters mentioned above were used as features for three ML models-KNN, LR, and SVM. The target variable in this case was the WEC forecast error. Given the relatively small size of our dataset (122 total events), we employed the leave-one-out cross-validation (LOOCV) technique to maximize the size of the test set. LOOCV works by iterating over the dataset, where for each iteration one event is ``left out'' and used as the test set, while the remaining 121 events are used to train the model. As a result, there were a total of 122 iterations. During each of them, the ML model was trained on 121 events and predictions for the ``left out'' event were recorded. This process was repeated until predictions were obtained for all events in the dataset. 

As described by equation \ref{eq:01}, the PE is defined as the difference between the ENLIL-predicted TOA and the observed TOA, expressed as \( \Delta t = t_{\text{enlil}} - t_{\text{obs}} \). The ML-predicted output is \( \Delta t_{\text{ml}}\), which represents learned error in ENLIL predictions based on errors in SW parameters. By subtracting this ML-predicted bias from the original ENLIL prediction error, we remove the systematic inaccuracies introduced by the SW errors.

\[
\Delta t - \Delta t_{\text{ml}} = t_{\text{enlil}} - t_{\text{obs}} - \Delta t_{\text{ml}} = t_{\text{ml\_enlil}} - t_{\text{obs}}.
\]

Here $t_{\text{ml\_enlil}}$ is the ''ML-corrected'' TOA from ENLIL. The expression, \( t_{\text{ml\_enlil}} - t_{\text{obs}} \), is the ``ML-corrected'' PE. To assess the performance of the ML models, we calculated the ``ML-corrected'' PE for each event.

\subsubsection{Univariate ML runs}

\begin{figure}[h!]
    \centering    \includegraphics[width=0.8\textwidth]{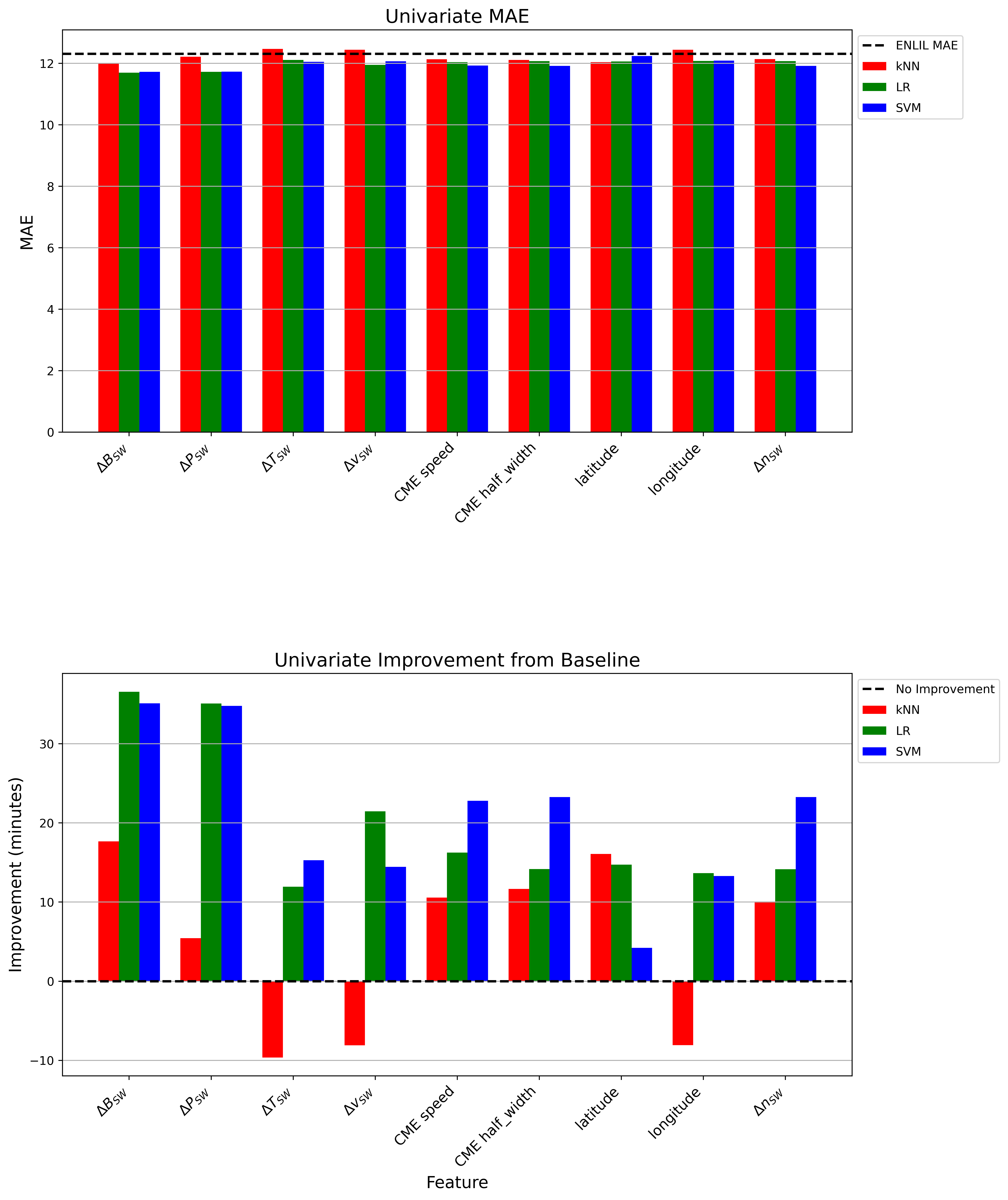}
    \caption{Comparison of univariate ML model performance with ENLIL. (a) The top panel represents MAE for univariate ML runs using different features. The red, green, and blue bars represent the ML-corrected forecast errors for the KNN, LR, and SVM models, respectively. For a comparison, the dashed line represents the ENLIL's MAE. (b) The bottom panel displays Improvement in arrival time prediction (in minutes) for univariate ML models relative to ENLIL. Positive values indicate improved accuracy over the baseline model.}
    \label{fig:uni_mae}
\end{figure}

A univariate ML model takes in only one feature as input to predict the target value. For example, a univariate model may take in \(\Delta B\) and predict the PE based solely on this feature. This prediction will not incorporate any information about \(\Delta P\), \(\Delta n\), \(\Delta T\), or \(\Delta v\). Figure~\ref{fig:uni_mae} and Table~\ref{tab:uni_mae} summarize the results of our univariate machine learning experiments. The top panel (of \ref{fig:uni_mae}) displays the corrected MAEs achieved by the ML models in hours, where the dashed line represents the original WEC MAE of 12.31. In contrast, the bottom panel illustrates the improvement in prediction accuracy (in minutes) resulting from various model-feature combinations. The red, green, and blue bars represent the ML-corrected forecast error results for the KNN, LR, and SVM models, respectively. Most of our ML runs successfully reduced the WEC forecast error. Out of 27 runs, only 3 did not result in error reduction, which demonstrates the robustness of the selected ML methods in minimizing forecast error.

\begin{table}[h]
\centering
\scriptsize 
\setlength{\tabcolsep}{6pt} 
\renewcommand{\arraystretch}{1.2} 
\begin{tabular}{|c|c|c|c|}
\hline
\textbf{Feature} & \textbf{KNN} & \textbf{SVM} & \textbf{LR} \\ \hline
$\Delta B_{SW}$   & 17.4  & 35.4  & 36.6  \\ \hline
$\Delta P_{SW}$   & 5.4   & 34.8  & 35.4  \\ \hline
$\Delta T_{SW}$   & -9.6  & 15.6  & 12.0  \\ \hline
$\Delta v_{SW}$   & -7.8  & 14.4  & 21.6  \\ \hline
CME speed         & 10.8  & 22.8  & 16.2  \\ \hline
CME half\_width   & 11.4  & 23.4  & 14.4  \\ \hline
Latitude          & 16.2  & 4.2   & 15.0  \\ \hline
Longitude         & -7.8  & 13.2  & 13.8  \\ \hline
$\Delta n_{SW}$   & 10.2  & 23.4  & 14.4  \\ \hline
\end{tabular}
\caption{MAE reduction (in minutes) for the 27 univariate model runs.}
\label{tab:uni_mae}
\end{table}

\begin{figure}[htbp]
    \centering
    \includegraphics[width=0.8\textwidth]{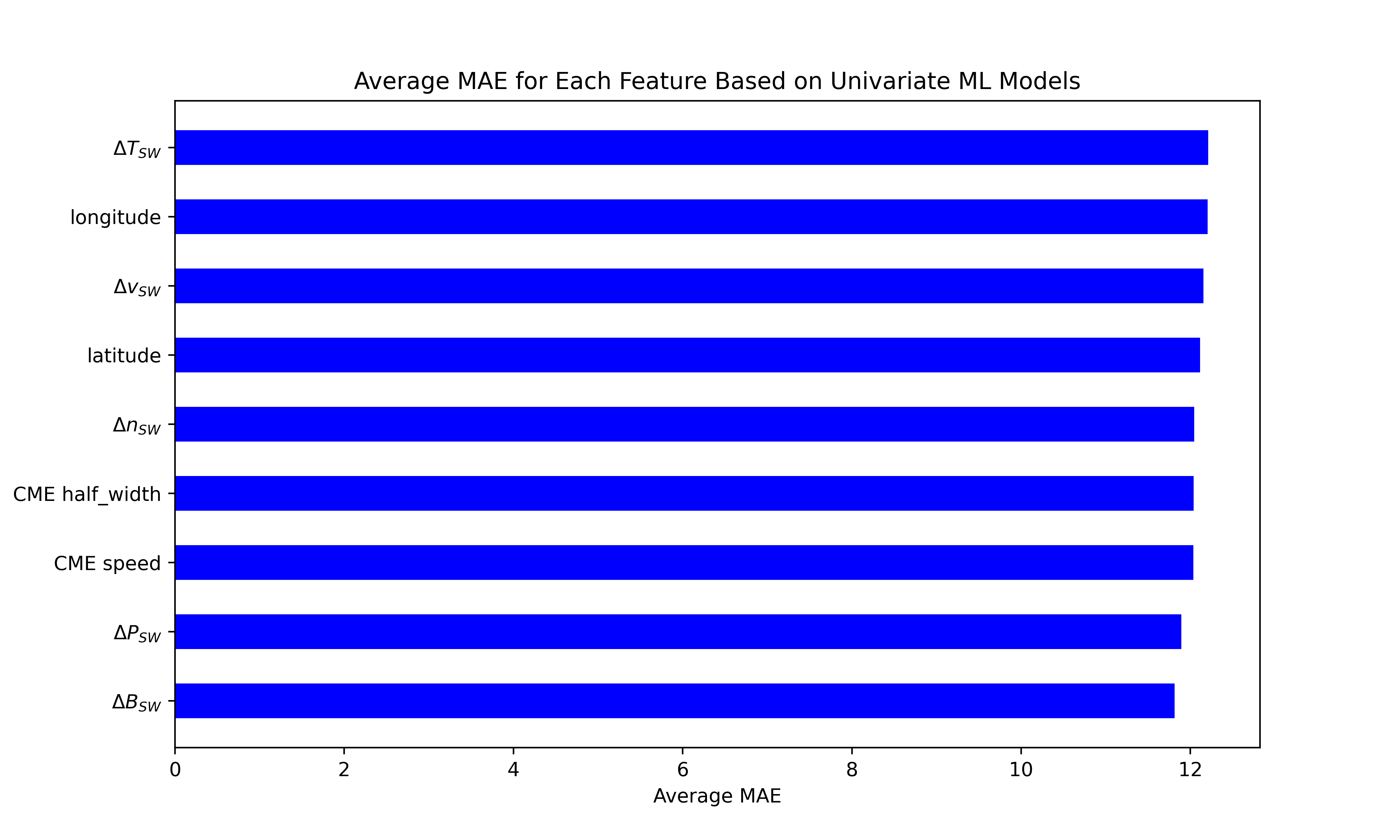}
    \caption{Rankings of features based on their average MAE scores across the KNN, LR, and SVM models. Features with lower MAE scores indicate higher predictive power.}
    \label{fig:ranked}
\end{figure}

Interestingly, all three instances for which the error was not reduced involved the KNN model, when it was applied to $\Delta T_{SW}$, $\Delta v_{SW}$, and the Cone model apex longitude. All instances of using the SVM and LR models reduced the forecast error. Among the tested features, it can be seen qualitatively that $\Delta B_{SW}$ and $\Delta P_{SW}$ are the best overall parameters in isolation. The most significant reduction was observed with the LR model applied to $\Delta B_{SW}$, achieving an error reduction to 11.70, closely followed by LR with $\Delta P_{SW}$, which reduced the error to 11.72. These correspond to reductions of 36.6 and 35.4 minutes, respectively. Overall, the univariate corrections lead to improvement 89\% of the time, with a mean gain of 18.25 minutes; in the remaining 11\%, predictions worsen by 8.4 minutes on average. Since a few features reduce MAE by over half an hour, it is tempting to apply a multivariate approach. A multivariate ML model can take multiple features into account simultaneously, potentially combining and amplifying their predictive power. However, before implementing a multivariate approach, it is critical to rank the features based on their predictive capabilities. This allows one to use the most impactful features selectively. The features are ranked based on the average of their MAE scores across all three ML models. These rankings are shown in Figure \ref{fig:ranked}. As expected, $\Delta B_{SW}$ and $\Delta P_{SW}$ emerged as the top-ranked features, having the lowest MAE scores. Interestingly, two of the Cone model parameters
-- the half-width and the CME initial speed -- ranked the third and fourth, respectively.

\begin{figure}[h!]
    \centering    \includegraphics[width=0.8\textwidth]{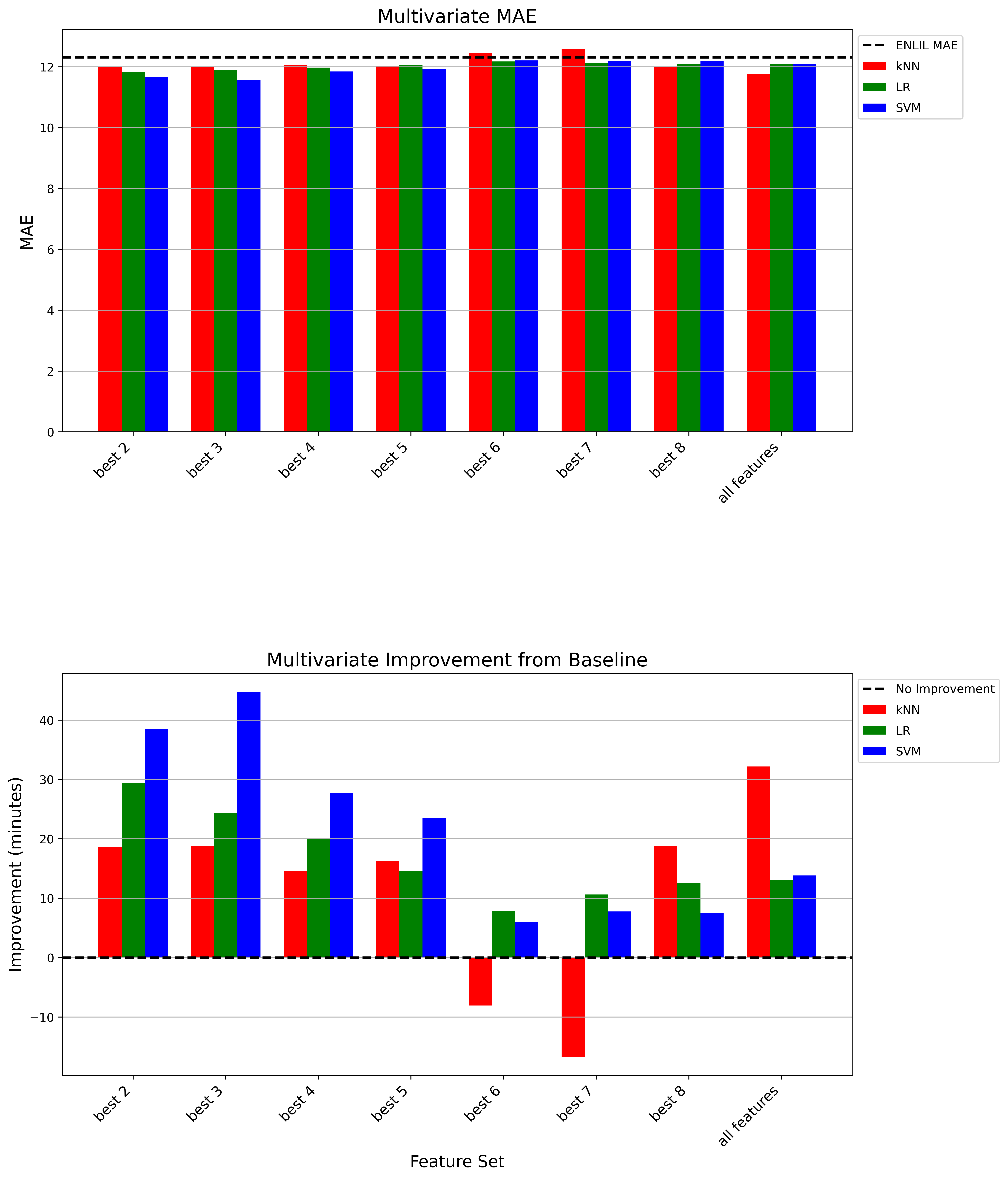}
    \caption{Comparison of multivariate ML model performance with ENLIL. The top panel represents MAE for the multivariate ML runs using different sets of features. The red, green, and blue bars represent the ML-corrected forecast error for the KNN, LR, and SVM models, respectively. For a comparison, the dashed line represents the ENLIL's MAE. The bottom panel displays Improvement in arrival time prediction (in minutes) for multivariate ML models relative to ENLIL. Positive values indicate improved accuracy over the baseline model.}
    \label{fig:multi_mae}
\end{figure}

\subsubsection{Multivariate ML runs}

Eight instances of the multivariate ML models were run for each of the three techniques (KNN, LR, and SVM), resulting in 24 runs in total. Based on the feature rankings shown in Figure \ref{fig:ranked}, the top two best features were used in the first instance, the top three best features were used in the second instance, and so on. These results are presented in Figure \ref{fig:multi_mae}. The improvements are also shown in table \ref{tab:multi_mae}. Out of these 24 runs, only two did not reduce the MAE below 12.31 hours. The exceptions were the KNN model with the best 6 and 7 parameters. The general trend observed is that when the first three or four best features were used as input to the multivariate models, they outperformed the univariate results. However, using the best 5 or more features does not improve the PE, as seen from the bars, which appear  to be quite close to the ENLIL MAE for the best 5-feature and any larger feature set. 
Using only the top two ranked features together with the SVM model, the MAE was reduced to 11.67 hours, which corresponds to a reduction by 39 minutes compared to the baseline. This is already better than the best performing univariate run (36.6 minutes). However, the best overall results were achieved by using the three top-ranked features: $\Delta B_{SW}$, $\Delta P_{SW}$, and CME initial speed with the SVM model. In this case, the MAE was reduced to 11.56 hours, representing the total reduction of 45 minutes. 

\begin{table}[h]
\centering
\scriptsize 
\setlength{\tabcolsep}{5pt} 
\renewcommand{\arraystretch}{1.1} 
\begin{tabular}{|c|c|c|c|}
\hline
\textbf{Feature Set} & \textbf{KNN} & \textbf{SVM} & \textbf{LR} \\ \hline
Best 2        & 18.6   & 39.0  & 29.4  \\ \hline
Best 3        & 18.6  & 45.0  & 24.6  \\ \hline
Best 4        & 15.0  & 27.6  & 20.4  \\ \hline
Best 5        & 16.2  & 23.4  & 14.4  \\ \hline
Best 6        & -7.8  & 6.0  & 8.4   \\ \hline
Best 7        & -16.8   & 7.8  & 10.8  \\ \hline
Best 8        & 19.2  & 7.8  & 12.6  \\ \hline
All Features  & 32.4  & 14.4  & 13.2  \\ \hline
\end{tabular}
\caption{MAE reduction (in minutes) for the multivariate models.}
\label{tab:multi_mae}
\end{table}

Overall, multivariate corrections lead to improvement 92\% of the time, with a mean gain of 19.30 minutes. For the remaining 8\%, prediction is worsen by 12.3 minutes. These improvements are skewed by the fact that the multivariate runs do not perform well with a large number of multiple features. For instance, when considering only the top-performing subsets, specifically the best 2, 3, 4, and 5 runs, the improvement rate increases to 100\%, with an average gain of 24.35 minutes.

\section{Discussion and Conclusions}
\label{discussion}

In this study, we analyze the impact of the inaccuracies in the description of the ambient SW, as obtained from the WEC model, on the CME forecast errors. A total of 122 CME events have been examined. Three ML models: (1) KNN, (2) LR, and (3) SVM have been used to quantify and correct the forecast errors caused specifically by the errors in reproducing the properties of the ambient SW at the L1 point. For the 122 events considered, the WEC model exhibited a tendency toward early predictions, with CMEs generally arriving a few hours later than the forecasted TOA. The MAE of CME TOA for these events was found to be 12.31 hours.

The goal of this article was to explore how the background SW computed with the WEC model can affect the CME TOAs depending on how different it is from the observed SW profile. Many studies have highlighted the uncertainties contained within the modeled SW \citep[e.g.][]{Owens2008, Riley2018}. In contrast with the initial parameters of CMEs, SW properties at L1 are directly measurable and can be used to improve CME forecasts. However, these observations are limited to a single point, L1, while CMEs are large, three-dimensional structures.  
To understand their interaction with the ambient SW, one would need observational data at multiple locations throughout the inner heliosphere. Nevertheless, we have demonstrated that discrepancies between L1 observations and simulations can still be quantified and used to improve CME forecasts. However, these improvement are not substantial. We have developed both the univariate and multivariate ML codes. The univariate model was used to rank the SW and Cone model parameters in terms of MAE reduction. As a result $\Delta B_{SW}$ and $\Delta P_{SW}$ emerged as the best performing parameters, while the CME Cone model speed and half width ranked third and fourth, respectively. This ranking of parameters was used to guide our multivariate runs. The MAE reduction of 45 minutes was found by using $\Delta B_{SW}$, $\Delta P_{SW}$, and CME initial speed with SVM model.

Although we use the CME Cone model parameters as features in our ML models, it is important to note that we are not correcting the errors associated with these parameters, as it is outside of the scope of this study. Unlike the SW forecast at L1 point, the true values for initial parameters of CMEs (i.e., speed, angular width, propagation direction) are not known. Other methods, such as ensemble modeling are used to quantify the initial parameters of CMEs (e.g \citet{Emmons2013}, where they used 100 sets of Cone model parameter to generate parameter distributions for ensemble modeling of 15 Earth-directed CMEs). \citet{Singh2023Improving} used a hybrid approach combining ML methods with ensemble modeling to quantify the errors in the initial CME parameters. The error bars form the Graudated Cylindrical Shell (GCS) model approximation of the initial CME parameters were used to generate ensemble members. The leading edge for an observed CME and ensemble members were tracked through the inner heliosphere using the observed and synthetic $j$-maps. They used 2 ML techniques (1) lasso regression and (2) neural network to correct the TOA guided by the ensemble member $j$-map distribution. They were able to reduce the overall MAE of the 6 events they studied from \(\sim 8\) hours to \(\sim 4\) hours. 

This study has several limitations that warrant consideration. One potential source of error in the CME TOA estimates may stem from the CME model employed. Specifically, the Cone model used in this study is not entirely realistic for CME propagation (\cite{Xie2004}), as it does not account for the CME flux-rope structure or its magnetic properties. Consequently, this simplification may fail to accurately capture the dynamic, magnetically driven interactions between the CME and the surrounding SW, potentially leading to erroneous TOA predictions. Therefore, using similar ML techniques trained on data from a flux rope CME model may provide different results. Another source of errors could be the DONKI catalog itself, which may contain inaccuracies. Since ML training relies on DONKI data, any deficiencies in data quality could impair the predictive correction capabilities of our models. 

In conclusion, reducing SW error at the L1 point using ML methods can yield an average correction of around 45 minutes. This suggests L1 measurements alone are insufficient to fully capture the complex interaction between CMEs and the background SW, which depends on the large-scale spatial distribution of plasma and magnetic fields upstream of Earth. To address this, we plan to conduct ensemble magnetohydrodynamic (MHD) simulations, exploring combinations of background SW and CME initial conditions to evaluate their respective impacts on TOA and identify which factor plays the dominant role in CME TOA forecast. Prior studies \citep[e.g.][]{Singh2023Improving, Mays15, Emmons2013} have highlighted the importance of constraining initial CME parameters through ensemble modeling and studying their importance in determining the CME TOA. Building on this, we plan to conduct ensemble MHD simulations of CMEs and constraining their propagation through the inner heliosphere using Heliospheric Imager data coupled with ML techniques.

\section{Conflict of Interest Statement}
The authors have no conflicts of interest to disclose.

\section{Acknowledgements}

The authors are grateful for the support provided jointly by the National Science Foundation (NSF) and NASA Space Weather with Quantified Uncertainties Program (NSF award AGS-2028154 and NASA grant 80NSSC20K1582). SR was supported by NASA FINNEST grant 80NSSC23K1636. TS was supported partially by NASA grant 80NSSC23K1592. Computing time allocations were also provided by NASA High-End Computing Program award SMD-17-1537 and ACCESS project MCA07S033.

\section{Open Research Section}

We acknowledge the Community Coordinated Modeling Center (CCMC) at GSFC for the DONKI database \citep{Maddox2014} used for CME data mining, \url{https://kauai.ccmc.gsfc.nasa.gov/DONKI/}. We also acknowledge the CCMC ISWA, \url{https://ccmc.gsfc.nasa.gov/tools/ISWA/} for the ENLIL SW files and the use of NASA/GSFC's Space Physics Data Facility's OMNIWeb (or CDAWeb or ftp) service, and OMNI data,  \cite{OMNI1hour2020}. The figures included in this document were generated using matplotlib \url{https://matplotlib.org/}. The Python code used to perform data preprocessing and ML tasks is archived on Zenodo and publicly available at \cite{Raza2025}.

\bibliography{Raza2024}
\end{document}